\documentstyle[twoside]{article}

\catcode`\@=11
\long\def\@makefntext#1{
\protect\noindent \hbox to 3.2pt {\hskip-.9pt
$^{{\eightrm\@thefnmark}}$\hfil}#1\hfill}               

\def\thefootnote{\fnsymbol{footnote}}
\def\@makefnmark{\hbox to 0pt{$^{\@thefnmark}$\hss}}    

\def\ps@myheadings{\let\@mkboth\@gobbletwo
\def\@oddhead{\hbox{}
\rightmark\hfil\eightrm\thepage}
\def\@oddfoot{}\def\@evenhead{\eightrm\thepage\hfil
\leftmark\hbox{}}\def\@evenfoot{}
\def\sectionmark##1{}\def\subsectionmark##1{}}

\oddsidemargin=\evensidemargin
\renewcommand{\thefootnote}{\fnsymbol{footnote}}

\newcounter{sectionc}\newcounter{subsectionc}\newcounter{subsubsectionc}
\renewcommand{\section}[1] {\vspace{12pt}\addtocounter{sectionc}{1}
\setcounter{subsectionc}{0}\setcounter{subsubsectionc}{0}\noindent
        {\tenbf\thesectionc. #1}\par\vspace{5pt}}
\renewcommand{\subsection}[1] {\vspace{12pt}\addtocounter{subsectionc}{1}
        \setcounter{subsubsectionc}{0}\noindent
        {\bf\thesectionc.\thesubsectionc. {\kern1pt \bfit #1}}\par\vspace{5pt}}
\renewcommand{\subsubsection}[1] {\vspace{12pt}\addtocounter{subsubsectionc}{1}
        \noindent{\tenrm\thesectionc.\thesubsectionc.\thesubsubsectionc.
        {\kern1pt \tenit #1}}\par\vspace{5pt}}
\newcommand{\nonumsection}[1] {\vspace{12pt}\noindent{\tenbf #1}
        \par\vspace{5pt}}

\newcounter{appendixc}
\newcounter{subappendixc}[appendixc]
\newcounter{subsubappendixc}[subappendixc]
\renewcommand{\thesubappendixc}{\Alph{appendixc}.\arabic{subappendixc}}
\renewcommand{\thesubsubappendixc}
        {\Alph{appendixc}.\arabic{subappendixc}.\arabic{subsubappendixc}}

\renewcommand{\appendix}[1] {\vspace{12pt}
        \refstepcounter{appendixc}
        \setcounter{figure}{0}
        \setcounter{table}{0}
        \setcounter{lemma}{0}
        \setcounter{theorem}{0}
        \setcounter{corollary}{0}
        \setcounter{definition}{0}
        \setcounter{equation}{0}
        \renewcommand{\thefigure}{\Alph{appendixc}.\arabic{figure}}
        \renewcommand{\thetable}{\Alph{appendixc}.\arabic{table}}
        \renewcommand{\theappendixc}{\Alph{appendixc}}
        \renewcommand{\thelemma}{\Alph{appendixc}.\arabic{lemma}}
        \renewcommand{\thetheorem}{\Alph{appendixc}.\arabic{theorem}}
        \renewcommand{\thedefinition}{\Alph{appendixc}.\arabic{definition}}
        \renewcommand{\thecorollary}{\Alph{appendixc}.\arabic{corollary}}
        \renewcommand{\theequation}{\Alph{appendixc}.\arabic{equation}}
        \noindent{\tenbf Appendix \theappendixc #1}\par\vspace{5pt}}
\newcommand{\subappendix}[1] {\vspace{12pt}
        \refstepcounter{subappendixc}
        \noindent{\bf Appendix \thesubappendixc. {\kern1pt \bfit #1}}
        \par\vspace{5pt}}
\newcommand{\subsubappendix}[1] {\vspace{12pt}
        \refstepcounter{subsubappendixc}
        \noindent{\rm Appendix \thesubsubappendixc. {\kern1pt \tenit #1}}
        \par\vspace{5pt}}

\newcommand{\textlineskip}{\baselineskip=13pt}
\newcommand{\smalllineskip}{\baselineskip=10pt}

\def\eightcirc{
\begin{picture}(0,0)
\put(4.4,1.8){\circle{6.5}}
\end{picture}}
\def\eightcopyright{\eightcirc\kern2.7pt\hbox{\eightrm c}}

\newcommand{\copyrightheading}[1]
        {\vspace*{-2.5cm}\smalllineskip{\flushleft
        {\footnotesize $\eightcopyright$\, World Scientific Publishing
         Company}\\
         }}

\def\abstracts#1#2#3{{
        \centering{\begin{minipage}{4.5in}\baselineskip=10pt\footnotesize
        \parindent=0pt #1\par
        \parindent=15pt #2\par
        \parindent=15pt #3
        \end{minipage}}\par}}

\newcommand{\bibit}{\nineit}
\newcommand{\bibbf}{\ninebf}
\renewenvironment{thebibliography}[1]
        {\frenchspacing
         \ninerm\baselineskip=11pt
         \begin{list}{\arabic{enumi}.}
        {\usecounter{enumi}\setlength{\parsep}{0pt}
         \setlength{\leftmargin 12.7pt}{\rightmargin 0pt} 
         \setlength{\itemsep}{0pt} \settowidth
        {\labelwidth}{#1.}\sloppy}}{\end{list}}

\newcounter{itemlistc}
\newcounter{romanlistc}
\newcounter{alphlistc}
\newcounter{arabiclistc}

\newcommand{\fcaption}[1]{
        \refstepcounter{figure}
        \setbox\@tempboxa = \hbox{\footnotesize Fig.~\thefigure. #1}
        \ifdim \wd\@tempboxa > 5in
           {\begin{center}
        \parbox{5in}{\footnotesize\smalllineskip Fig.~\thefigure. #1}
            \end{center}}
        \else
             {\begin{center}
             {\footnotesize Fig.~\thefigure. #1}
              \end{center}}
        \fi}

\newcommand{\tcaption}[1]{
        \refstepcounter{table}
        \setbox\@tempboxa = \hbox{\footnotesize Table~\thetable. #1}
        \ifdim \wd\@tempboxa > 5in
           {\begin{center}
        \parbox{5in}{\footnotesize\smalllineskip Table~\thetable. #1}
            \end{center}}
        \else
             {\begin{center}
             {\footnotesize Table~\thetable. #1}
              \end{center}}
        \fi}

\def\@citex[#1]#2{\if@filesw\immediate\write\@auxout
        {\string\citation{#2}}\fi
\def\@citea{}\@cite{\@for\@citeb:=#2\do
        {\@citea\def\@citea{,}\@ifundefined
        {b@\@citeb}{{\bf ?}\@warning
        {Citation `\@citeb' on page \thepage \space undefined}}
        {\csname b@\@citeb\endcsname}}}{#1}}

\newif\if@cghi
\def\cite{\@cghitrue\@ifnextchar [{\@tempswatrue
        \@citex}{\@tempswafalse\@citex[]}}
\def\citelow{\@cghifalse\@ifnextchar [{\@tempswatrue
        \@citex}{\@tempswafalse\@citex[]}}
\def\@cite#1#2{{$\null^{#1}$\if@tempswa\typeout
        {IJCGA warning: optional citation argument
        ignored: `#2'} \fi}}

\def\pmb#1{\setbox0=\hbox{#1}
        \kern-.025em\copy0\kern-\wd0
        \kern.05em\copy0\kern-\wd0
        \kern-.025em\raise.0433em\box0}

\def\fnt#1#2{\footnotetext{\kern-.3em
        {$^{\mbox{\scriptsize #1}}$}{#2}}}

\def\fpage#1{\begingroup
\voffset=.3in
\thispagestyle{empty}\begin{table}[b]\centerline{\footnotesize #1}
        \end{table}\endgroup}

\def\runninghead#1#2{\pagestyle{myheadings}
\markboth{{\protect\footnotesize\it{\quad #1}}\hfill}
{\hfill{\protect\footnotesize\it{#2\quad}}}}
\font\tenrm=cmr10
\font\tenit=cmti10
\font\tenbf=cmbx10
\font\bfit=cmbxti10 at 10pt
\font\ninerm=cmr9
\font\nineit=cmti9
\font\ninebf=cmbx9
\font\eightrm=cmr8

\def\qed{\hbox{${\vcenter{\vbox{                        
   \hrule height 0.4pt\hbox{\vrule width 0.4pt height 6pt
   \kern5pt\vrule width 0.4pt}\hrule height 0.4pt}}}$}}

\renewcommand{\thefootnote}{\fnsymbol{footnote}}        

\def\bsc{{\sc a\kern-6.4pt\sc a\kern-6.4pt\sc a}}       
\def\bflatex{\bf L\kern-.30em\raise.3ex\hbox{\bsc}\kern-.14em
T\kern-.1667em\lower.7ex\hbox{E}\kern-.125em X}

\begin{document}

\runninghead{{\tt WBase}: a C package to reduce tensor products $\ldots$}
{{\tt WBase}: a C package to reduce tensor products $\ldots$}

\normalsize\textlineskip
\thispagestyle{empty}
\setcounter{page}{1}

\copyrightheading{}                     

\vspace*{0.88truein}

\fpage{1}
\centerline{\bf {\tt WBase}: a C package to reduce tensor products}
\vspace*{0.035truein}
\centerline{\bf of Lie algebra representations. Description and new
developments.\footnote{Supported
in part by M.P.I. This work is carried out in the framework of
the European Community Programme ``Gauge Theories, Applied Supersymmetry
and Quantum Gravity'' with a financial contribution under contract
SC1-CT92-D789.}}
\vspace*{0.37truein}
\centerline{\footnotesize Antonio Candiello}
\vspace*{0.015truein}
\centerline{\footnotesize\it Dipartimento di Fisica, Universit\`a di Padova}
\baselineskip=10pt
\centerline{\footnotesize\it Istituto Nazionale di Fisica Nucleare,
Sezione di Padova}
\baselineskip=10pt
\centerline{\footnotesize\it Padova, 35131, Italy}
\vspace*{0.225truein}

\vspace*{0.21truein}
\abstracts{A non trivial application of a modern computer language ("C")
in a highly structured and object-oriented fashion is presented.
The contest is that of Lie algebra representations (irreps), specifically
the problem of reducing the products of irreps with the weight tree
algorithm. The new {\tt WBase 2.0} version with table-generation and Young
tableaux display capabilities is introduced.}{}{}



\vspace*{1pt}\textlineskip      

\def\bea{\begin{eqnarray}}      \def\eea{\end{eqnarray}}
\def\beq{\begin{equation}}      \def\eeq{\end{equation}}

\section{Introduction}    
\vspace*{-0.5pt}
\noindent
Calculations in algebra representation theory, in particular decompositions of
products of irreps, are needed in several sectors of physics.
The Dynkin approach to the representation theory\cite{SLANSKY}
is known by physicists thanks to its generality: all simple Lie algebras,
included the exceptional ones, are described and manipulated in the same
formal environment. In the Dynkin approach the
algebras are described uniquely by the $l\times l$ Cartan matrix, where $l$
is the rank of the algebra. The irreps of a given algebra are identified by
a unique {\it highest weight vector\/} of $l$ positive integers.

The purpouse of this contribution is to show the convenience of using
modern computer programming techniques when applied to the Dynkin approach of
algebra representation theory. Indeed, we have been able to construct a
versatile algebra-manipulation package, named {\tt WBase},\cite{CANDIELLO} such
that: 1) {\tt WBase} is a compact project, and the memory the Dynkin approach
needs is used at best so that it works also on small computers;
2) {\tt WBase} is easily upgradable with features.

With regard to the point 2) in this work we will describe the new features
of the {\tt WBase V2.0} version, a) the new table-generation routines;
and b) the new Young-display support routines. The {\tt WBase V2.0}
now supports directly all Cartan algebras, both classical and
exceptional.

Algorithms more specialized but faster than the Dynkin one use the
extended Young diagrams. In {\tt WBase V2.0} we added the extended Young
diagram display capability for the classical algebras in order to analize
these alternative methods.

\pagebreak
\textheight=7.8truein
\setcounter{footnote}{0}
\renewcommand{\thefootnote}{\alph{footnote}}

\section{Dynkin's approach to representation theory\cite{SLANSKY}}\noindent
A simple Lie algebra in the Cartan-Weyl basis
is described by a set of $l$ simultaneously diagonalizable generators $H_i$
and by the other generators $E_\alpha$, satisfying
\beq
[H_i,H_j]=0\qquad i=1,\ldots,l
\eeq
\beq
[H_i,E_\alpha]=\alpha_iE_\alpha,\qquad i=1,\ldots,l;\
\alpha=-\frac{d-l}{2},\ldots,\frac{d-l}{2};
\eeq
$l$ is the {\it rank} of the algebra, and the set of $l$-vectors
$\alpha_i$ are the {\it roots}. It results that all roots can be constructed
via linear combinations by a set of $l$ roots, called {\it simple roots}.

By the simple roots one then constructs the $l\times l$ {\it Cartan matrix},
which is the key to the classification of the Lie
algebras, and it is known for all of them: the $A_n$, $B_n$, $C_n$, $D_n$
series and the exceptional algebras $G_2$, $F_4$, $E_6$, $E_7$, $E_8$.
In the Dynkin approach the Cartan matrix is all we need to
completely describe the algebra. This is at the basis of
our package: the routine {\tt wstartup}, given the name of
the algebra, takes care of generating
algorithmically the related Cartan matrix (called {\tt wcart} in {\tt WBase}).

The {\it metric } $G_{ij}$, which is related to the inverse of the Cartan
matrix (called {\tt wmetr} in {\tt WBase}) introduces a scalar product in
the space of {\it weight vectors}, which are $l$-uples of integer numbers.
Each irrep in an algebra is uniquely classified by a weight vector, the {\it
highest weight\/} $\Lambda$ whose components are all positive integers
(the Dynkin labels). The different states in an irrep in a given irrep are
again described by a weight vector $w$; the full set of all states of a
given irrep is thus described by a set of weight vectors, called the {\it
weight system\/}.

The dimension of an irrep $\Lambda$ can be calculated with the help of the {\it
Weyl formula} (encoded in the {\tt weyl} function),
\beq
\dim(\Lambda)=\prod_{pos. roots \alpha}
\frac{(\Lambda+\delta,\alpha)}{(\delta,\alpha)}
\eeq
where $\Lambda$ is the highest weight determining the irrep, $\delta=
(1,\cdots,1)$, and $(\,,\,)$ is the scalar product constructed with the metric
$G_{ij}$. The {\it positive roots} are the positive weight vectors of the
weight system of the adjoint irrep.

The weight system of an irrep is needed to reduce the products of irreps, which
is one of the main task of our package. The weight system is obtained by
subtracting the simple roots $\alpha_i$ from the highest weight as described by
the following recursive procedure (handled by {\tt wtree}):
\vskip .5truecm
{\em
\hskip.5truecm for all $i$ in $1,\ldots,l$\par
\hskip.9truecm compute $u_k=\Lambda-k\alpha_i$, $k=1,\cdots,u(i)$\par
\hskip.9truecm add all $u_k$ to the weight system\par
\hskip.9truecm restart the procedure with $\Lambda:=u_k$}
\vskip .5truecm
The computational heaviest part of the construction of the weight system
is the computation of the degeneration of each weight vector. It is computed
by the {\it Freudenthal recursion formula} (encoded in {\tt freud}),
which needs the degeneration of previous levels,
\beq
\hbox{deg}(w)=\frac{
 2\sum_{pos. roots \alpha}\sum_{k>0}
 \hbox{deg}(w+k\alpha)\,(w+k\alpha,\alpha)
 }{\parallel\Lambda+\delta\parallel^2-\parallel w+\delta\parallel^2}
\eeq
with the initial condition $\hbox{deg}(\Lambda)=1$.

Once developed the machinery to generate full weight systems along with their
degenerations, we can reduce the products of irreps, i.e. solve for the
$\Lambda_i$ in the equation
\beq
\Lambda_a\otimes\Lambda_b=\Lambda_1\oplus\cdots\oplus\Lambda_n,
\eeq
each with its degeneration, where each highest weights $\Lambda$ identify
uniquely an irrep. The algorithm (encoded in {\tt wpdisp}) is the following.
{\tt wsyst}: generate the weight systems $ws(\Lambda_a)=\{w_{an}\}$,
$ws(\Lambda_b)=\{w_{bm}\}$, and ({\tt bprod}): construct the set
$P=\{w_{an}+w_{bm}\}$; then, until $P$ is empty,
({\tt whighest}): find the highest weight $\Lambda_i$ in $P$,
({\tt wsyst}): generate the weight system
$ws(\Lambda_i)=\{w_{ik}\}$, and {(\tt bremove}): subtract from $P$
each weight vector in $ws(\Lambda_i)$.

\section{Dynamic data structures}\noindent
The fundamental data type underlying our routines is the dynamically
allocated (by {\tt walloc}/{\tt wfree}) weight vector {\tt wvect}, with a
run-time chosen length ${\tt wsize}\equiv l$.
The extended weight vector {\tt wplace} contains also the degeneration and
level informations of the weight vectors embedded in the weight systems.
The {\tt wplace} data type is used essentially to give a structure to the raw
data kept in the blocks on which we base our granularity-flexible allocation
scheme for the weight systems.

The difficulty in constructing such weight systems for arbitrary irreps is due
to the fact that the dimension of the table needed to store the weight systems
is not known in advance. The only solution that does not waste large amounts
of memory and does not limit our routines more than the hardware does is a
multiple block allocation scheme. The data structure is as follows:

\medskip

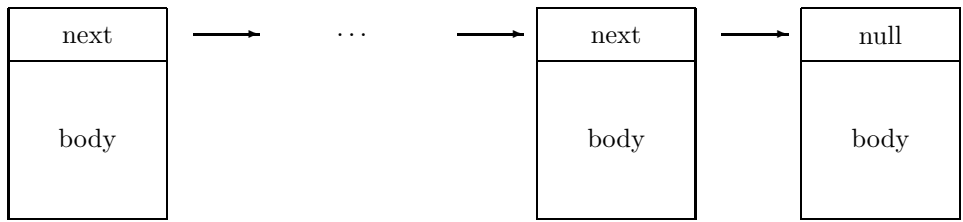
\begin{picture}(500,100)
\put(0,100){\line(1,0){60}}
\put(0,80){\line(1,0){60}}
\put(0,20){\line(1,0){60}}
\put(0,20){\line(0,1){80}}
\put(60,20){\line(0,1){80}}
\put(30,90){\makebox(0,0){next}}
\put(30,50){\makebox(0,0){body}}
\put(70,90){\vector(1,0){25}}

\put(100,90){\makebox(60,0){$\ldots$}}
\put(170,90){\vector(1,0){25}}

\put(200,100){\line(1,0){60}}
\put(200,80){\line(1,0){60}}
\put(200,20){\line(1,0){60}}
\put(200,20){\line(0,1){80}}
\put(260,20){\line(0,1){80}}
\put(230,90){\makebox(0,0){next}}
\put(230,50){\makebox(0,0){body}}
\put(270,90){\vector(1,0){25}}

\put(300,100){\line(1,0){60}}
\put(300,80){\line(1,0){60}}
\put(300,20){\line(1,0){60}}
\put(300,20){\line(0,1){80}}
\put(360,20){\line(0,1){80}}
\put(330,90){\makebox(0,0){null}}
\put(330,50){\makebox(0,0){body}}
\end{picture}

\noindent
a singly linked list of blocks of identical size fixed at run-time (according
to the fragmentation required) by the number of {\tt wplace} vector entries as
given in {\tt bsize}; the typedef that defines a single block leaves therefore
its main structure undefined, as it is different for different ranks {\tt
wsize} and block dimension {\tt bsize}:
\begin{verbatim}
typedef struct tblock {
 struct tblock *next;
 char body[1];} wblock;
\end{verbatim}
This definition, as the {\tt wplace} one, is used through casts that give
form to unstructured raw data as returned by the allocator {\tt balloc}.
Single blocks are released with a call to {\tt bfree}, linked blocks are
released with a call to {\tt bsfree} for the first of them.
The structured form of the blocks, when casting with {\tt wplace} and defining
{\tt bsize} and {\tt wsize}, is as follows:
\medskip

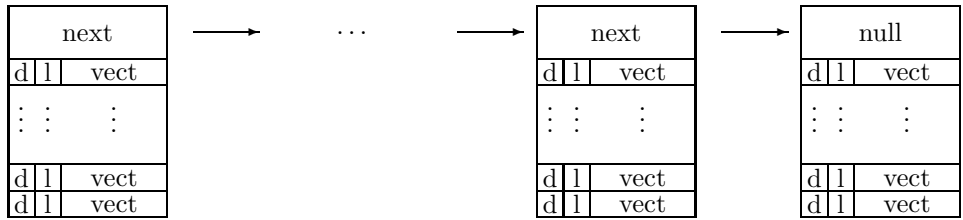
\begin{picture}(500,100)
\put(0,100){\line(1,0){60}}
\put(0,80){\line(1,0){60}}
\put(0,20){\line(1,0){60}}
\put(0,20){\line(0,1){80}}
\put(60,20){\line(0,1){80}}
\put(30,90){\makebox(0,0){next}}
\put(0,70){\line(1,0){60}}
\put(0,40){\line(1,0){60}}
\put(0,30){\line(1,0){60}}
\put(10,70){\line(0,1){10}}
\put(20,70){\line(0,1){10}}
\put(10,30){\line(0,1){10}}
\put(20,30){\line(0,1){10}}
\put(10,20){\line(0,1){10}}
\put(20,20){\line(0,1){10}}
\put(5,75){\makebox(0,0){d}}
\put(15,75){\makebox(0,0){l}}
\put(40,75){\makebox(0,0){vect}}
\put(5,35){\makebox(0,0){d}}
\put(15,35){\makebox(0,0){l}}
\put(40,35){\makebox(0,0){vect}}
\put(5,25){\makebox(0,0){d}}
\put(15,25){\makebox(0,0){l}}
\put(40,25){\makebox(0,0){vect}}
\put(5,60){\makebox(0,0){$\vdots$}}
\put(15,60){\makebox(0,0){$\vdots$}}
\put(40,60){\makebox(0,0){$\vdots$}}
\put(70,90){\vector(1,0){25}}

\put(100,90){\makebox(60,0){$\ldots$}}
\put(170,90){\vector(1,0){25}}

\put(200,100){\line(1,0){60}}
\put(200,80){\line(1,0){60}}
\put(200,20){\line(1,0){60}}
\put(200,20){\line(0,1){80}}
\put(260,20){\line(0,1){80}}
\put(230,90){\makebox(0,0){next}}
\put(200,70){\line(1,0){60}}
\put(200,40){\line(1,0){60}}
\put(200,30){\line(1,0){60}}
\put(210,70){\line(0,1){10}}
\put(220,70){\line(0,1){10}}
\put(210,30){\line(0,1){10}}
\put(220,30){\line(0,1){10}}
\put(210,20){\line(0,1){10}}
\put(220,20){\line(0,1){10}}
\put(205,75){\makebox(0,0){d}}
\put(215,75){\makebox(0,0){l}}
\put(240,75){\makebox(0,0){vect}}
\put(205,35){\makebox(0,0){d}}
\put(215,35){\makebox(0,0){l}}
\put(240,35){\makebox(0,0){vect}}
\put(205,25){\makebox(0,0){d}}
\put(215,25){\makebox(0,0){l}}
\put(240,25){\makebox(0,0){vect}}
\put(205,60){\makebox(0,0){$\vdots$}}
\put(215,60){\makebox(0,0){$\vdots$}}
\put(240,60){\makebox(0,0){$\vdots$}}
\put(270,90){\vector(1,0){25}}

\put(300,100){\line(1,0){60}}
\put(300,80){\line(1,0){60}}
\put(300,20){\line(1,0){60}}
\put(300,20){\line(0,1){80}}
\put(360,20){\line(0,1){80}}
\put(330,90){\makebox(0,0){null}}
\put(300,70){\line(1,0){60}}
\put(300,40){\line(1,0){60}}
\put(300,30){\line(1,0){60}}
\put(310,70){\line(0,1){10}}
\put(320,70){\line(0,1){10}}
\put(310,30){\line(0,1){10}}
\put(320,30){\line(0,1){10}}
\put(310,20){\line(0,1){10}}
\put(320,20){\line(0,1){10}}
\put(305,75){\makebox(0,0){d}}
\put(315,75){\makebox(0,0){l}}
\put(340,75){\makebox(0,0){vect}}
\put(305,35){\makebox(0,0){d}}
\put(315,35){\makebox(0,0){l}}
\put(340,35){\makebox(0,0){vect}}
\put(305,25){\makebox(0,0){d}}
\put(315,25){\makebox(0,0){l}}
\put(340,25){\makebox(0,0){vect}}
\put(305,60){\makebox(0,0){$\vdots$}}
\put(315,60){\makebox(0,0){$\vdots$}}
\put(340,60){\makebox(0,0){$\vdots$}}
\end{picture}

\section{High level routines}\noindent
The task of setting up the algebra structures and handle the lists of vectors
is taken by the following procedures:

{\tt wstartup(type,rank)}, {\tt wcleanup(string)}\par\noindent
These are the open/close routines for setting up the algebra; given e.g. the
algebra B5 (SO(11)) one would call {\tt wstartup('B',5)} to allocate the
related vectors and matrices, and then {\tt wcleanup()} to deallocate them.
Each invocation of {\tt wstartup()} must be followed by {\tt wcleanup()}.
Multiple invocations of this pair of routines are needed for
the construction of a table of several algebras.

{\tt wread(hw)}, {\tt wfdisp(hw)}, {\tt wydisp(hw)}\par\noindent
These are the input/output routines. {\tt wread} read a highest weight
vector label from the standard input. {\tt wfdisp(hw)} calls the functions
{\tt weyl}, {\tt casimir} and {\tt wheight} to display the related irrep
informations on the standard output. {\tt wydisp} is the new {\tt WBase V2.0}
extended Young tableaux display function available for all non exceptional
algebra types. These routines need a valid {\tt wvect} as handled by
{\tt walloc()}/{\tt wfree()}.

{\tt wsave(w,base,level)}, {\tt wremove(w,base)}\par\noindent
Insert/remove the weight vector {\tt w} in the list pointed by {\tt base}.
If {\tt level} is greater than zero, {\tt wsave} does a sorted
insertion without degeneration increment, otherwise {\tt w} is stored with
its degeneration. {\tt wremove} will remove a weight vector {\tt w} from the
list {\tt base} only if the degeneration is just 1.

{\tt wsyst(hw)}, {\tt bdisp(base)}, {\tt bsfree(base)}\par\noindent
{\tt wsyst} returns the full linked list of blocks of the weight system
of highest weight {\tt hw} along with the degenerations. It uses the
recursive algorithm described in section 2, by calling {\tt wsave} to store the
weight vectors, then computes the degeneration with the Freudenthal formula.
{\tt bdisp(base)} displays to the standard output all entries of
the weight system in {\tt base}. The block list constructed by
{\tt wsyst} has to be deallocated with {\tt bsfree}.

{\tt wpdisp(hw1,hw2,mod)}\par\noindent
This function hides all the complexities of reducing products of irreps and of
the underlying data structure, by giving to the standard output all the irreps
in the product, from the highest to the lowest, according to the modality
choosen by {\tt mod}\footnote{See the header file {\tt wbase.h}.}. The product
routines are also available as iteration functions (see below).

{\it User interface in {\tt WBase V2.0}}\par\noindent
In the file {\tt wmain.c} it is implemented an ANSI C terminal-like interface
with the user. A more sophisticated user interaction may be constructed taking
this file as an example. Thanks to the new capabilities introduced in
{\tt WBase V2.0}, we had to add more options which are still one-letter options
(for the details, refer directly to the source code).

\section{Iterators}\noindent
One of the standard conceptual device of the object-oriented technology
is the {\it iterator}. The iterator hides the implementation details
providing a consistent interface for moving through the data structure of
the object considered. In {\tt WBase} we introduced:
1) the iterator needed to scan the {\tt wblock} list which contains the
weight tree and 2) the iterator used to generate the {\tt wblock} lists which
contain the decomposed irreps of a product. New in {\tt WBase V2.0} is
3) the table-generation iterator. Following is a short description of their
use.

{\it Scanning through the {\tt wblock} list}\par\noindent
It is done through the {\tt pstart}/{\tt pnext} iterators:
\begin{verbatim}
wplace *p;
pcurr pc;
for(p=pstart(base,&pc);p!=NULL;p=pnext(&pc))
  < do something with p >
\end{verbatim}
remembering that: {\tt p->vect} gives the weight vector, {\tt p->deg} its
degeneration, and {\tt p->level} the level of the vector within the weight
system. To remove the last entry from the list one uses
{\tt base=plast(p,base)}, with the just removed vector returned in the area
pointed by {\tt p}.

{\it Getting the irreps of the decomposition of products}\par\noindent
The iteration functions {\tt wpstart(base)}/{\tt wpnext(base,b)} that
interface the construction of products return a
{\tt wblock} pointer to the full weight system of the reduced irrep.
In order to display the weight tree of each irrep in the product of
the two irreps with highest weight {\tt hw1} and {\tt hw2} the fragment of
code is as follows:
\begin{verbatim}
wblock *base,*b;
base=bprod(wsyst(hw1),wsyst(hw2));
for(b=wpstart(base);b!=NULL;b=wpnext(b,base))
  bdisp(b);
\end{verbatim}

{\it Generating irreps of increasing dimensions}\par\noindent
This is one of the main innovations of {\tt WBase V2.0} that were only
announced when we first introduced the package \cite{CANDIELLO}. The
table-generation procedure {\tt wsequence} is used to produce increasing
dimension irreps:
\begin{verbatim}
wblock *b=NULL;
wvect hw=walloc();
int dim;
while(b=wsequence(hw,b,maxdim,&dim))
 wfdisp(hw);
\end{verbatim}
{\tt wsequence} provides to: 1) allocate the first block when
invoked with b=NULL the first time, 2) allocate eventual subsequent blocks,
to store the encountered highest weights {\tt hw}, and 3) deallocate
all of them when {\tt maxdim} is reached. In this last case it returns
NULL. Like the other iteration procedures, {\tt wsequence} has as an
argument a specific iterator object (the {\tt wblock} pointer {\tt b})
which identify each irrep sequencing. It is then possible to nidificate
multiple iterators to produce table of products.

In the next version of {\tt WBase} we will probably transform the
algebra initialization/destruction functions {\tt wstartup}/{\tt wcleanup}
in iterators in order to extend the table-generation capabilities of
{\tt WBase V2.0} to multiple algebra tables. We will also probably switch to
the C++ language to provide a more consistent iteration interface across the
different data structures. The operator overloading capabilities of the C++
language will also simplify the interface of our package by unifying our
list storage/remove functions (by overloading the {\tt +=} and {\tt -=}
operators) and our input/output functions (by overloading the {\tt <<} and
{\tt >>} operators).

\section{Conclusions}\noindent
The {\tt WBase} package, in our opinion, represents a useful and self-contained
demonstration of the convenience of new object-oriented software technology
when combined with the C powerful dynamic data allocation facilities. As a
by-product, we obtained what we think can be a useful tool for not too heavy
irrep-related necessities of physicists. The Dynkin approach of Lie algebra
representation theory helped us to maintain a unified and elegant structure to
our package, however it must be noted that there are less general but faster
algorithms\cite{WYBOURNE} based on tensor and spinor manipulations useful
in computing products of irreps.

\nonumsection{References}
\end{document}